\documentclass{article}
\usepackage{spconf,amsmath,graphicx,booktabs,makecell}

\title{Transferring voice knowledge for Acoustic event detection: An empirical study}
\name{\begin{tabular}{c}
Dawei Liang$^{\star}$\sthanks{Work performed at Facebook.},
Yangyang Shi$^{\dagger}$, Yun Wang$^{\dagger}$, Nayan Singhal$^{\dagger}$, Alex Xiao$^{\dagger}$, \\
Jonathan Shaw$^{\dagger}$, Edison Thomaz$^{\star}$, Ozlem Kalinli$^{\dagger}$, Mike Seltzer$^{\dagger}$
\end{tabular}}
  
\address{$^{\star}$ University of Texas at Austin, USA \\
$^{\dagger}$ Facebook AI, USA}
\begin{document}
%
\maketitle

\begin{abstract}
Detection of common events and scenes from audio is useful for extracting and understanding human contexts in daily life. Prior studies have shown that leveraging knowledge from a relevant domain is beneficial for a target acoustic event detection (AED) process. Inspired by the observation that many human-centered acoustic events in daily life involve voice elements, this paper investigates the potential of transferring high-level voice representations extracted from a public speaker dataset to enrich an AED pipeline. Towards this end, we develop a dual-branch neural network architecture for the joint learning of voice and acoustic features during an AED process and conduct thorough empirical studies to examine the performance on the public AudioSet \cite{gemmeke2017audio} with different types of inputs. Our main observations are that: 1) Joint learning of audio and voice inputs improves the AED performance (mean average precision) for both a CNN baseline (0.292 vs 0.134 mAP) and a TALNet \cite{wang2019comparison} baseline (0.361 vs 0.351 mAP); 2) Augmenting the extra voice features is critical to maximize the model performance with dual inputs.

\end{abstract}

\vspace{-3pt}
\section{Introduction}
\vspace{-3pt}
\label{sec:intro}

With the development of modern smart devices with listening capabilities such as voice assistants, smartphones, and wearable devices, audio has been increasingly used as a modality for the inference of human activities, and contexts \cite{lane2015deepear, liang2019audio}. Acoustic event detection (AED) is the process of detecting the type and temporal onset/offset of acoustic events within an audio stream. While existing AED models have been advanced for modeling the target audio, studies have shown that knowledge transfer from a relevant domain is beneficial to boost the learning process further and the model capabilities \cite{aytar2016soundnet, kong2020panns}. For example, knowledge transfer is especially useful when data in the target domain is not sufficient for model generalization \cite{tan2018survey}, which is a typical case when dealing with real-world audio. 

Many acoustic event types captured in daily life are related to human voice or contain voice elements, such as conversations, TV/radio sounds, music, or sounds from a crowd. To the best of our knowledge, however, very few prior attempts have studied the opportunities of transferring and incorporating voice knowledge into a general AED process. This paper investigates this opportunity by jointly training audio and high-level voice representations for an AED model based on dual-branch neural network architecture. We observe the benefits of adding extra voice inputs to a convolutional neural network (CNN) and a TALNet \cite{wang2019comparison} baseline by using AudioSet \cite{gemmeke2017audio} as the test dataset. Specifically, our study demonstrates a few strategies that bridge the learning gap between the audio and the voice features.

\vspace{-3pt}
\section{Related work}
\vspace{-3pt}
\label{sec:related}

The general process of AED is to build a classification model where the existence of an acoustic class is determined by the output class probability from the model. Conventional approaches include the usage of statistical models based on hand-crafted features \cite{vuegen2013mfcc, uzkent2012non}. Recent work has increasingly focused on using neural networks for the modeling of audio \cite{gorin2016dcase, espi2015exploiting, wang2016audio, parascandolo2016recurrent} given their success in computer vision. 

Due to the increasing scale of audio data exposure, modern audio datasets are typically weak-labeled by annotators. The dataset only gives the label for the whole recording without detailed frame-level annotation. However, in practice, frame-level labeling is required. The following work \cite{wang2019comparison, yu2018multi, kong2018audio, chou2018learning} has addressed the issue. The TALNet \cite{wang2019comparison} is one of the state-of-the-art efforts for AED with weakly labeled audio inputs, which has demonstrated strong performance for acoustic event tagging and localization at the same time. 

Transferring knowledge from a source domain to a target task can be a useful way to enrich the learning of the target dataset \cite{tan2018survey}. It is particularly meaningful in real-world audio analysis where the target audio accessibility can be limited due to challenges such as scalability \cite{chon2013understanding} and privacy constraints \cite{liang2020characterizing}. For acoustic classification, transfer learning has been successfully applied both across tasks \cite{kong2019cross, dissanayake2020speech,niz2020,tonami2019joint, zhang2019cross} and across modalities \cite{aytar2016soundnet, kaya2017video}. Specifically, the extraction and leveraging of pre-trained neural network embeddings is a common way of audio knowledge transfer \cite{wang2019comparison, kong2019cross, hershey2017cnn}. Compared to conventional hand-crafted voice features such as i-vectors or the mel features \cite{zhao2013analyzing, schmidt1996speaker, schmidt2014large, al2017comparison}, voice embeddings are directly obtained from a neural network trained for speaker voice classification. The voice embedding represents the knowledge that the network has learned to identify the speaker patterns \cite{snyder2018x, lukic2017learning, wan2018generalized, nagrani2017voxceleb}. Our study aims to leverage voice embeddings extracted from an existing speaker dataset to enrich the AED process. As far as we know, this is the first effort to incorporate knowledge from voice inputs for AED on the AudioSet corpus.

\vspace{-3pt}
\section{Architecture}
\vspace{-3pt}
\label{sec:Architecture}

\vspace{-3pt}
\subsection{Overall pipeline}
\vspace{-3pt}
\label{ssec:pipeline}

\begin{figure}[t]
\centerline{\includegraphics[width=9cm]{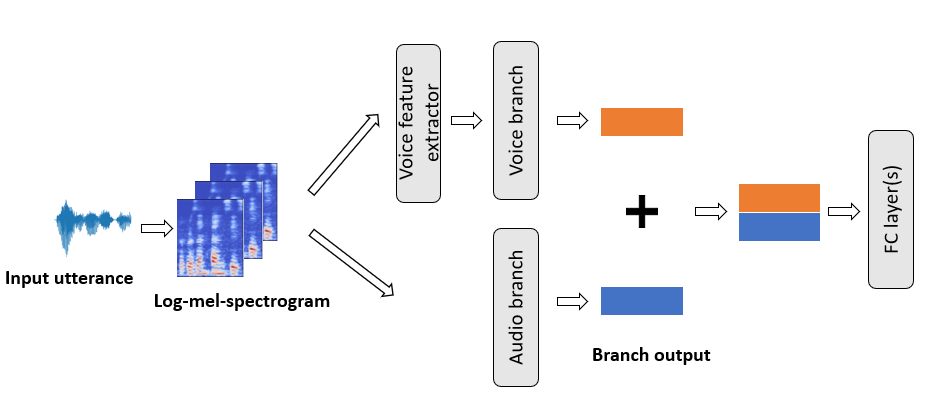}}

%
%
\caption{Overall pipeline of our study. The acoustic classifier is a neural network with two input branches, taking as input the log-mel features and extracted voice embeddings of an utterance respectively.}
\label{fig:overall}
\vspace{-4pt}
\end{figure}

Fig. \ref{fig:overall} shows the overall pipeline of our study. The pipeline consists of two steps - feature extraction and acoustic event detection. The first step extracts the log-mel features of an audio input utterance. In addition to leveraging the log-mel features as the AED inputs, we apply an extra pre-trained model to extract voice embedding representations from the audio.  Applying voice embedding transfers pre-trained voice knowledge of the feature extractor to the target audio, and it applies to both vocal and non-vocal input. In the second step, the acoustic classifier is a neural network architecture with two input branches: an audio branch with log-mel inputs and a voice branch with the generated voice feature inputs. The outputs of both branches are concatenated along the feature dimension and fed to the final fully connected layer(s). We did not apply early fusion of the features, because feature fusion at an intermediate layer left us more flexibility to optimize the two input branches separately \cite{khaleghi2013multisensor}. In our study, the voice feature extractor was trained with an existing speaker dataset. Once pre-trained, the parameters of the feature extractor were fixed, and only the audio and voice branches were trained for the target AED task.

\vspace{-3pt}
\subsection{Audio branch architecture}
\vspace{-3pt}
\label{ssec:pipeline}
The audio branch was developed for the log-mel inputs. In our study, we started with a shallow CNN baseline and then the TALNet. The CNN architecture is as follows:

\textbf{Input $\rightarrow$ Conv1[64] $\rightarrow$ Conv2[128] $\rightarrow$ Conv3[256] $\rightarrow$ Conv4[256] $\rightarrow$ FC[2048] $\rightarrow$ FC[1024] $\rightarrow$ FC[527]}

\noindent where ConvX[$K$] denotes a 2D convolutional layer with the ReLU \cite{agarap2018deep} activation and $K$ channels. The kernel size, padding, and stride were (3$\times$3), (1$\times$1), and (1$\times$1). Besides, a max-pooling of size (2$\times$2), (2$\times$2), and (1$\times$2) was added for Conv1, Conv2 and Conv4, respectively. FC[$K$] denotes a fully connected layer of size $K$ with the ReLU activation. We adopted the model architectures for both baselines, excluding their fully connected layer(s) as our audio branch. We then added back the fully connected layer(s) after feature fusion.

\vspace{-3pt}
\subsection{Voice branch architecture}
\vspace{-3pt}
\label{ssec:pipeline}

\begin{figure}[t]
\centerline{\includegraphics[width=9cm]{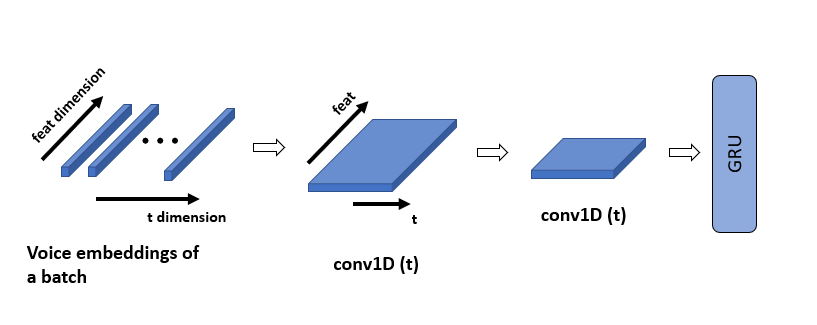}}

\caption{Architecture of the voice branch. It consists of 1D convolutional layers on the time (t) dimension followed by a uni-directional GRU layer.}
\vspace{-6pt}
\label{fig:voice_branch}
\end{figure}

Unlike the audio branch, we did not apply convolution on the feature dimension of the voice inputs since adjacent elements of an embedding may not have spatial correlation as the log-mel vectors do. Hence, our voice branch consists of 1D convolutional layers along the temporal dimension of the voice embeddings. The feature dimension of the embeddings is mapped to the channel dimension of the convolutional layers. Fig. \ref{fig:voice_branch} shows such a process. In such a design, reducing the number of network channels of each convolutional layer essentially reduces the size of the feature dimension. We added an extra uni-directional GRU layer following the convolutional layers to improve the learning performance.


\vspace{-3pt}
\section{Voice representations}
\vspace{-3pt}
\label{sec:representations}

\vspace{-4pt}
\subsection{Pre-training of voice models}
\vspace{-4pt}
\label{ssec:Dataset}

To develop the voice feature extractor, we constructed a speaker recognition task where a network was trained to classify given speaker voice classes. The task was built on the public VoxCeleb1 \cite{nagrani2017voxceleb} speaker dataset. The dataset consists of audio utterances of over 1K celebrities from public YouTube videos of varying lengths. Specifically, we leveraged 1,211 speakers in the dataset for our model training and validation. For each speaker, ten utterances were randomly selected for model validation, and the rest were used for training, resulting in an average of 109 utterances per speaker in our training set.

The audio utterances were sampled at 16kHz and truncated or padded to 10 seconds. We then extracted 64D log-mel features using a frame length of 64ms and a frame shift of 25ms. The resulting log-mel features of a minibatch of input to our voice models had a shape of (batch$\times$1$\times$400$\times$64). The features were normalized per dimension by the mean and standard deviation calculated over the entire training set. Two voice models were then applied, inspired from \cite{nagrani2017voxceleb} (\textit{Arch1}) and a simplified version of TALNet (\textit{Arch2}). We slightly modified the model parameters to fit our requirements, as shown in Table \ref{speaker}. The models were deployed using PyTorch \cite{NEURIPS2019_9015}. ReLU activation and batch normalization were applied except for the last fc layer. Besides, the ``ceil'' mode was enabled for the max-pooling layers. For Arch1, the output of the last conv2D layer was re-shaped from (batch$\times$1024$\times$100$\times$1) to (batch$\times$100$\times$1024). The temporal dimension (100) was then aggregated by average pooling. For Arch2, the temporal dimension was aggregated after the fc layer.

\begin{table}[t]%
\begin{center}
\begin{tabular}{cc}
  \toprule
  \small \textbf{Arch1} & \small \textbf{Arch2} \\
  \midrule
    \small conv2D (96, 3$\times$3, 1, 1) & \small conv2D (32, 3$\times$3, 1, 1)  \\
    \small mpool (2$\times$2)  & \small mpool (2$\times$2) \\
    \small conv2D (256, 3$\times$3, 1, 1)  & \small conv2D (32, 3$\times$3, 1, 1) \\
    
    \small mpool (2$\times$2)  & \small conv2D (64, 3$\times$3, 1, 1) \\
    \small conv2D (384, 3$\times$3, 1, 1)  & \small mpool (2$\times$2) \\
    \small conv2D$\times$2 (256, 3$\times$3, 1, 1) & \small conv2D (64, 3$\times$3, 1, 1) \\
    \small mpool (1$\times$2)  & \small flatten $\rightarrow$ (batch$\times$100$\times$1024) \\
    \small conv2D (1024, 1$\times$8, 1, 0)  &
    \small biGRU (512$\times$2) \\
    \small fc$\times$2 (1024 / 1211)  & \small fc (1211) \\
  \bottomrule
\end{tabular}
\caption{Architecture of our voice models. conv2D: 2D convolutional layers (channel, kernel size, stride, padding); mpool: max pooling; fc: fully-connected layers.}
\vspace{-12pt}
\label{speaker}

\end{center}

\end{table}%

To set up training, we used an initial learning rate of 2$\times10^{-4}$ and a batch size of 25. We applied the same class balancing strategy as in \cite{wang2019comparison} to account for class imbalance. The learning rate was shrunk by a factor of 0.9 when the validation accuracy plateaued, and training was stopped when the learning rate reached 1$\times10^{-6}$. We used the Adam \cite{kingma2014adam} optimizer and the cross-entropy loss.

\vspace{-4pt}
\subsection{Voice feature extraction}
\vspace{-4pt}
\label{ssec:Dataset}

The best validation accuracy we obtained was 95.8\% and 93.4\% for Arch1 and Arch2, respectively. The high accuracy values were as expected because of the large utterance size we used. The voice embeddings were then extracted before average pooling for Arch1 and following the GRU layer for Arch2, which maintains the temporal information of the input utterance at a resolution of 10Hz. The resulting embeddings were then of shape (batch$\times$100$\times$1024) for a batch. In the following sections, we will refer to the embeddings from the two models as \textit{emb\_1} and \textit{emb\_2} for convenience.

\vspace{-4pt}
\section{Experiments}
\vspace{-4pt}
\label{sec:Experiments}

\vspace{-4pt}
\subsection{Training setup}
\vspace{-4pt}
\label{ssec:setup}

We leveraged AudioSet for our AED study. The dataset consists of over 2 million 10-second audio utterances of 527 annotated acoustic classes, including vocal and non-vocal sounds extracted from public YouTube videos. We used the same evaluation set containing 24,832 utterances.

We followed our speaker recognition steps to derive the same type of log-mel features as inputs to the AED part. We then tested the two baselines of the audio branch independently. In deployment, we removed Conv4 of the CNN baseline for our dual-branch tests to maintain a similar model size with and without voice inputs. Besides, we used a hidden size of 768 for the GRU layer of TALNet. For both baselines, we enabled the ``ceil'' mode of the max-pooling layers in PyTorch. The output of the audio branch was consistently in a shape of (batch$\times$100$\times$768), where the temporal size was 100 (0.1s resolution).

For the voice branch, we applied two 1D convolutional layers so that the parameter size of the voice branch could be less considerable ($<$1M) compared to the baseline models. The kernel size, padding, and stride were 3, 1, and 1, respectively, with ReLU activation. Batch normalization was also added before the activation. In the convolutional layers, the number of channels was 256 and 64, respectively.  For the GRU layer, the hidden size was 64. Hence, the outputs of the voice branch were of shape (batch$\times$100$\times$64). 

The final shape of features was (batch$\times$100$\times$832) after fusion. The last fully connected layer was of size 527 with the sigmoid activation. The per-class binary predictions were aggregated for an utterance by linear softmax pooling on the frame-level probability outputs. We used the same learning rate and optimization setup for training as in the speaker recognition task, but the validation metric was switched to the mean average precision (mAP) score which AudioSet also used. Besides, we used the binary cross-entropy loss.

\vspace{-4pt}
\subsection{Experiments and result discussions}
\vspace{-4pt}
\label{ssec:setup}

We first examined the maximum performance of the models with augmented voice inputs. Inspired by common augmentation strategies for acoustic features, we processed the voice embeddings with three strategies -- time masking \cite{park2019specaugment}, mixup \cite{zhang2017mixup}, and adding dropout \cite{srivastava2014dropout} to the voice branch. Specifically, we randomly masked 40 voice embedding frames out of the total 100 for each input utterance. The mixup was applied by mixing up labels and the corresponding input features at a batch level. We applied this for both branches at the same time with an alpha value of 1. Besides, we dropped out the voice features with a probability of 0.5.

\begin{table}[t]%
\begin{center}
\begin{tabular}{cccc}
  \toprule
  \small \textbf{Combination} & \small \textbf{mAP} & \small \textbf{mAUC} & \small \textbf{\textit{d}-prime} \\
  \midrule
    \small CNN         & \small 0.134 & \small 0.903 & \small 1.840 \\
    \small CNN+emb1    & \small 0.292 & \small 0.951 & \small 2.338 \\
    \small CNN+emb2    & \small 0.256 & \small 0.950 & \small 2.325 \\

    \small TALNet      & \small 0.351 & \small 0.966 & \small 2.584 \\
    \small TALNet+emb1 & \small 0.360 & \small 0.962 & \small 2.506 \\
    \small TALNet+emb2 & \small 0.361 & \small 0.962 & \small 2.517 \\
  \bottomrule
\end{tabular}
\caption{Overall AED results with different input types and base architectures of the audio branch.}
\vspace{-12pt}
\label{raw}

\end{center}

\end{table}%

Table \ref{raw} shows the overall AED results based on combinations of the input types and the audio branch architectures. In addition to the best mAP values, we also reported the best mean area under the curve (mAUC) and $d$-prime metrics to show the performance better. The results with both baselines show that adding the extra voice branch improves the AED performance regarding the mAP metric, and the observation was consistent for both types of voice inputs. As expected, the performance jump was much bigger for the CNN baseline than for TALNet, since TALNet is a more sophisticated model for the AudioSet AED task, and the supplementary information carried by the voice inputs can be relatively marginal. Interestingly, the mAUC and $d$-prime metrics degraded for TALNet with the voice inputs, possibly because the training was only optimized for the validation mAP. We also tried doubling the number of convolutional layers for the voice branch (with 1/2 of the channel size per layer to maintain the same output shape). However, no improvement was observed for either type of embeddings (0.357 mAP for both emb1 and emb2), indicating that 2 layers of CNN were sufficient for learning from the high-level voice features.

We then studied the AED effects with and without augmentation applied to the voice inputs. Table \ref{aug} shows the best validation epochs, the corresponding loss values and the mAP for the test cases. From the comparisons, it can be clearly seen that adding augmentation is critical to improve joint training of the two branches and maximize the model performance. Table \ref{case} further demonstrates the effects of individual types of augmenting strategies with the TALNet-based audio branch. While all techniques improved the joint training process, we can see that mixup augmentation yielded the biggest improvement. For a better comparison, we also ran a test for the original TALNet baseline with the same setup of mixup augmentation on the log-mel inputs. However, no improvement was observed in our test (0.342 mAP), indicating that the augmentation process contributed more significantly with the extra voice inputs than with the original baseline model. 

By checking the class-wise predictions of the TALNet-based models, we found that the predicting performance led most ($\ge$ 0.06 mAP) with the voice embeddings compared to the baseline model for the 10 classes: \textit{Hoot, Dental\_drill, Yodeling, Air\_horn, Gobble, Roar, Ringtone, Television, Snoring, Video\_game\_music}. An interesting finding was that adding the extra voice inputs did not guarantee a better prediction for the voice-related AudioSet classes. For example, the mAP for \textit{Speech} was 0.776 / 0.778 / 0.782 with the original TALNet / TALNet+emb1 / TALNet+emb2, but the values were 0.320 / 0.314 / 0.336 for \textit{Singing} and 0.209 / 0.186 / 0.206 for \textit{Conversation}. A possible explanation is that the training of the two branches was an integrated process and the models were optimized for the global performance, sometimes at a compromise of such individual classes.

\begin{table}[t]%
\begin{center}
\begin{tabular}{lclclcl}
  \toprule
  \small \textbf{Combination} & \small \textbf{Epoch} & \small \textbf{Train / Val loss} & \small \textbf{mAP}\\
  \midrule
    \small CNN+emb1, no aug & \small 11 & \small 9.7 / 13.3 & \small 0.229\\
    \small CNN+emb1, with aug  & \small 39 & \small 12.7 / 12.2 & \small 0.264 \\
    \small CNN+emb2, no aug  & \small 13 & \small 8.9 / 12.8 & \small 0.256\\
    \small CNN+emb2, with aug  & \small 45 & \small 12.6 / 11.7 & \small 0.292 \\
    \midrule
    \small TALNet+emb1, no aug  & \small 25 & \small 6.7 / 11.8 & \small 0.328 \\
    \small TALNet+emb1, with aug  & \small 56 & \small 11.4 / 10.6 & \small 0.360 \\
    \small TALNet+emb2, no aug  & \small 19 & \small 7.6 / 11.5 & \small 0.331 \\
    \small TALNet+emb2, with aug  & \small 71 & \small 11.4 / 10.6 & \small 0.361 \\
  \bottomrule
\end{tabular}
\caption{Results with and without augmentation on the voice features. Feature augmentation is a critical factor for better joint training of the two input branches.}
\vspace{-6pt}
\label{aug}

\end{center}

\end{table}%

\begin{table}[t]%
\begin{center}
\begin{tabular}{lclclcl}
  \toprule
  \small \textbf{Strategy} & \small \textbf{Epoch} & \small \textbf{Train / Val loss} & \small \textbf{mAP}\\
  \midrule
    \small emb1+dp  & \small 21 & \small 7.1 / 11.4 & \small 0.334 \\
    \small emb1+tmask  & \small 25 & \small 6.8 / 11.7 & \small 0.331 \\
    \small emb1+mixup  & \small 62 & \small 10.6 / 10.8 & \small 0.350 \\
    \midrule
    \small emb2+dp  & \small 20 & \small 7.3 / 11.4 & \small 0.334 \\
    \small emb2+tmask  & \small 19 & \small 7.8 / 11.5 & \small 0.334 \\
    \small emb2+mixup  & \small 68 & \small 11.6 / 10.7 & \small 0.356 \\
  \bottomrule
\end{tabular}
\caption{Results with a single type of augmenting strategies. \textit{dp}: dropout; \textit{tmask}: time-mask augmentation; \textit{mixup}: mixup augmentation. Mixup augmentation is the most effective approach to improve the training.}
\vspace{-12pt}
\label{case}

\end{center}

\end{table}%





\vspace{-6pt}
\section{Conclusions}
\vspace{-6pt}

\label{sec:Experiments}

This paper explored a novel approach for acoustic event detection by incorporating pre-trained voice embeddings into an AED pipeline. Towards this end, we developed a dual-branch neural network architecture for joint training of the inputs. We then reported the overall and class-wise performance with a CNN baseline and a strong TALNet baseline developed on AudioSet. Our results showed the benefits of adding extra voice inputs to the tested models (0.292 vs 0.134 mAP for the CNN baseline and 0.361 vs 0.351 mAP for TALNet baseline). Furthermore, we showed that adding augmentation and dropout on the voice inputs is critical to maximize the model performance with dual inputs.



\small \bibliographystyle{IEEEbib}
\small \bibliography{ref}

\begin{thebibliography}{10}

\bibitem{gemmeke2017audio}
Jort~F Gemmeke et~al.,
\newblock ``Audio set: An ontology and human-labeled dataset for audio
  events,''
\newblock in {\em ICASSP}, 2017.

\bibitem{wang2019comparison}
Yun Wang, Juncheng Li, and Florian Metze,
\newblock ``A comparison of five multiple instance learning pooling functions
  for sound event detection with weak labeling,''
\newblock in {\em ICASSP}, 2019.

\bibitem{lane2015deepear}
Nicholas~D Lane, Petko Georgiev, and Lorena Qendro,
\newblock ``Deepear: robust smartphone audio sensing in unconstrained acoustic
  environments using deep learning,''
\newblock in {\em UbiComp}, 2015.

\bibitem{liang2019audio}
Dawei Liang and Edison Thomaz,
\newblock ``Audio-based activities of daily living (adl) recognition with
  large-scale acoustic embeddings from online videos,''
\newblock {\em IMWUT}, 2019.

\bibitem{aytar2016soundnet}
Yusuf Aytar, Carl Vondrick, et~al.,
\newblock ``Soundnet: Learning sound representations from unlabeled video,''
\newblock {\em NIPS}, 2016.

\bibitem{kong2020panns}
Qiuqiang Kong, Yin Cao, et~al.,
\newblock ``Panns: Large-scale pretrained audio neural networks for audio
  pattern recognition,''
\newblock {\em IEEE/ACM Transactions on Audio, Speech, and Language
  Processing}, vol. 28, pp. 2880--2894, 2020.

\bibitem{tan2018survey}
Chuanqi Tan et~al.,
\newblock ``A survey on deep transfer learning,''
\newblock in {\em ICANN}, 2018.

\bibitem{vuegen2013mfcc}
Lode Vuegen, BVD Broeck, et~al.,
\newblock ``An mfcc-gmm approach for event detection and classification,''
\newblock in {\em WASPAA}, 2013.

\bibitem{uzkent2012non}
Burak Uzkent, Buket~D Barkana, et~al.,
\newblock ``Non-speech environmental sound classification using svms with a new
  set of features,''
\newblock {\em International Journal of Innovative Computing, Information and
  Control}, vol. 8, no. 5, pp. 3511--3524, 2012.

\bibitem{gorin2016dcase}
Arseniy Gorin, Nurtas Makhazhanov, and Nickolay Shmyrev,
\newblock ``Dcase 2016 sound event detection system based on convolutional
  neural network,''
\newblock {\em IEEE AASP Challenge: Detection and Classification of Acoustic
  Scenes and Events}, 2016.

\bibitem{espi2015exploiting}
Miquel Espi et~al.,
\newblock ``Exploiting spectro-temporal locality in deep learning based
  acoustic event detection,''
\newblock {\em EURASIP Journal on Audio, Speech, and Music Processing}, 2015.

\bibitem{wang2016audio}
Yun Wang, Leonardo Neves, and Florian Metze,
\newblock ``Audio-based multimedia event detection using deep recurrent neural
  networks,''
\newblock in {\em ICASSP}. IEEE, 2016, pp. 2742--2746.

\bibitem{parascandolo2016recurrent}
Giambattista Parascandolo, Heikki Huttunen, and Tuomas Virtanen,
\newblock ``Recurrent neural networks for polyphonic sound event detection in
  real life recordings,''
\newblock in {\em ICASSP}, 2016.

\bibitem{yu2018multi}
Changsong Yu et~al.,
\newblock ``Multi-level attention model for weakly supervised audio
  classification,''
\newblock {\em arXiv:1803.02353}, 2018.

\bibitem{kong2018audio}
Qiuqiang Kong et~al.,
\newblock ``Audio set classification with attention model: A probabilistic
  perspective,''
\newblock in {\em ICASSP}, 2018.

\bibitem{chou2018learning}
Szu-Yu Chou, Jyh-Shing~Roger Jang, and Yi-Hsuan Yang,
\newblock ``Learning to recognize transient sound events using attentional
  supervision.,''
\newblock in {\em IJCAI}, 2018.

\bibitem{chon2013understanding}
Yohan Chon et~al.,
\newblock ``Understanding the coverage and scalability of place-centric
  crowdsensing,''
\newblock in {\em UbiComp}, 2013.

\bibitem{liang2020characterizing}
Dawei Liang, Wenting Song, and Edison Thomaz,
\newblock ``Characterizing the effect of audio degradation on privacy
  perception and inference performance in audio-based human activity
  recognition,''
\newblock in {\em MobileHCI}, 2020.

\bibitem{kong2019cross}
Qiuqiang Kong, Yin Cao, et~al.,
\newblock ``Cross-task learning for audio tagging, sound event detection and
  spatial localization: Dcase 2019 baseline systems,''
\newblock {\em arXiv:1904.03476}, 2019.

\bibitem{dissanayake2020speech}
Vipula Dissanayake et~al.,
\newblock ``Speech emotion recognition'in the wild'using an autoencoder.,''
\newblock in {\em INTERSPEECH}, 2020.

\bibitem{niz2020}
Niko Moritz, Gordon Wichern, et~al.,
\newblock ``All-in-one transformer: Unifying speech recognition, audio tagging,
  and event detection,''
\newblock {\em Interspeech}, 2020.

\bibitem{tonami2019joint}
Noriyuki Tonami et~al.,
\newblock ``Joint analysis of acoustic events and scenes based on multitask
  learning,''
\newblock in {\em IEEE WASPAA}, 2019.

\bibitem{zhang2019cross}
Ruixiong Zhang, Wei Zou, and Xiangang Li,
\newblock ``Cross-task pre-training for on-device acoustic scene
  classification,''
\newblock {\em arXiv:1910.09935}, 2019.

\bibitem{kaya2017video}
Heysem Kaya, Furkan G{\"u}rp{\i}nar, and Albert~Ali Salah,
\newblock ``Video-based emotion recognition in the wild using deep transfer
  learning and score fusion,''
\newblock {\em Image and Vision Computing}, 2017.

\bibitem{hershey2017cnn}
Shawn Hershey et~al.,
\newblock ``Cnn architectures for large-scale audio classification,''
\newblock in {\em icassp}, 2017.

\bibitem{zhao2013analyzing}
Xiaojia Zhao and DeLiang Wang,
\newblock ``Analyzing noise robustness of mfcc and gfcc features in speaker
  identification,''
\newblock in {\em ICASSP}, 2013, pp. 7204--7208.

\bibitem{schmidt1996speaker}
Michael Schmidt and Herbert Gish,
\newblock ``Speaker identification via support vector classifiers,''
\newblock in {\em ICASSP}, 1996.

\bibitem{schmidt2014large}
Ludwig Schmidt, Matthew Sharifi, and Ignacio~Lopez Moreno,
\newblock ``Large-scale speaker identification,''
\newblock in {\em ICASSP}, 2014.

\bibitem{al2017comparison}
Musab~TS Al-Kaltakchi, Wai~L Woo, et~al.,
\newblock ``Comparison of i-vector and gmm-ubm approaches to speaker
  identification with timit and nist 2008 databases in challenging
  environments,''
\newblock in {\em EUSIPCO}, 2017.

\bibitem{snyder2018x}
David Snyder et~al.,
\newblock ``X-vectors: Robust dnn embeddings for speaker recognition,''
\newblock in {\em ICASSP}, 2018.

\bibitem{lukic2017learning}
Yanick~X Lukic et~al.,
\newblock ``Learning embeddings for speaker clustering based on voice
  equality,''
\newblock in {\em IEEE MLSP}, 2017.

\bibitem{wan2018generalized}
Li~Wan et~al.,
\newblock ``Generalized end-to-end loss for speaker verification,''
\newblock in {\em ICASSP}. IEEE, 2018.

\bibitem{nagrani2017voxceleb}
Arsha Nagrani, Joon~Son Chung, and Andrew Zisserman,
\newblock ``Voxceleb: a large-scale speaker identification dataset,''
\newblock {\em arXiv:1706.08612}, 2017.

\bibitem{khaleghi2013multisensor}
Bahador Khaleghi et~al.,
\newblock ``Multisensor data fusion: A review of the state-of-the-art,''
\newblock {\em Information fusion}, 2013.

\bibitem{agarap2018deep}
Abien~Fred Agarap,
\newblock ``Deep learning using rectified linear units (relu),''
\newblock {\em arXiv:1803.08375}, 2018.

\bibitem{NEURIPS2019_9015}
Adam Paszke et~al.,
\newblock ``Pytorch: An imperative style, high-performance deep learning
  library,''
\newblock in {\em NIPS}. 2019.

\bibitem{kingma2014adam}
Diederik~P Kingma and Jimmy Ba,
\newblock ``Adam: A method for stochastic optimization,''
\newblock {\em arXiv:1412.6980}, 2014.

\bibitem{park2019specaugment}
Daniel~S Park et~al.,
\newblock ``Specaugment: A simple data augmentation method for automatic speech
  recognition,''
\newblock {\em arXiv:1904.08779}, 2019.

\bibitem{zhang2017mixup}
Hongyi Zhang et~al.,
\newblock ``mixup: Beyond empirical risk minimization,''
\newblock {\em arXiv:1710.09412}, 2017.

\bibitem{srivastava2014dropout}
Nitish Srivastava et~al.,
\newblock ``Dropout: a simple way to prevent neural networks from
  overfitting,''
\newblock {\em The journal of machine learning research}, 2014.

\end{thebibliography}

\end{document}